\documentclass[preprint]{article}
\usepackage{authblk}
\usepackage{graphicx}
\usepackage{dcolumn}
\usepackage{bm}
\usepackage{amsmath}
\usepackage{amssymb}
\usepackage{multirow}
\usepackage{caption}
\usepackage{subcaption}
\usepackage{epstopdf}
\usepackage{hyperref}

\title{\bf Observational viability of fractional holographic dark energy in LRS Bianchi type-I cosmological model }
\author[1]{Shivangi Rathore\thanks{shivangirathore1912@gmail.com}}
\author[2]{Elangbam Chingkheinganba Meetei\thanks{chingelang@gmail.com}}
\author[2]{S. Surendra Singh\thanks{ssuren.mu@gmail.com}}

\affil[1,2]{Department of Mathematics, National Institute of Technology Manipur, Imphal-795004,India.}

\begin{document}

\maketitle

\textbf{Abstract}:

We scrutinized the Locally Rotational Symmetry (LRS) Bianchi type-I cosmological model in the presence of fractional holographic dark energy.  We calculate the value of the Hubble parameter $H(z)$ using the field equations. After that, we fit the model by employing the MCMC techniques to observational data, which includes Hubble, Hubble+BAO, Hubble+Pantheon+Shoes, and Hubble+BAO+Pantheon+Shoes datasets. With the help of these datasets, we calculate the model parameters, viz. $H_{0}(z), \alpha,$ and $\Omega_{m_{0}}$. The value of $H_{0}(z),\alpha$ and $\Omega_{m_{0}}$ lies in the range $67.688^{+1.246}_{-1.197}- 67.80^{+1.23}_{-1.23}, 0.86^{+0.14}_{-0.15}- 0.885^{+0141}_{-0.149}$ and $0.264^{+0.018}_{-0.018}- 0.27^{+0.02}_{-0.02}$ respectively. Our results indicate that the evolution of the density parameters corresponding to dark energy (DE) and dark matter (DM), particularly for $\alpha=1.01$, along with the transition at $z_{t} = 0.55$ of the deceleration parameter $q$ from positive values to $-1$, reflects a phase of accelerated expansion that closely resembles the $\Lambda$CDM model. The behavior of the equation of state (EoS) parameter further demonstrates that the Universe's evolution aligns well with the framework of the fractional holographic dark energy (FHDE) model, suggesting its compatibility with scenarios of late-time cosmology. Moreover, the analysis of the statefinder diagnostics ${(r,s)}$ and the present value of ${(r,s)} = (0.74,0.07)$ reveal characteristics that converge towards the $\Lambda$CDM fixed point. A comparison between the observational constraints on our model parameters and those of the $\Lambda$CDM model exhibits a strong degree of agreement, thereby reinforcing the physical plausibility and consistency of the proposed cosmological model.\\

\textbf{Keywords:} LRS Bianchi type-I cosmological model, Fractional holographic dark energy, Observational analysis.\\

\section{Introduction}\label{sec1}

Recent decades have seen extensive scientific inquiry into the Universe's enigmatic behavior, spanning phenomena such as early-time inflation, late-time acceleration, black holes, wormholes, dark energy, and gravitational waves. These investigations have probed the fundamental nature of the Universe. The cosmological principle, which assumes the Universe is isotropic and homogeneous on large scales, has been a cornerstone of understanding. However, the 1992 Cosmic Background Explorer discovery of small anisotropies in the cosmic microwave background radiation challenged this assumption, revealing a more complex Universe \cite{1}. These finding were further corroborated by subsequent measurements from experiments like Wilkinson Microwave Anisotropy Probe (WMAP) \cite{2}, Cosmic Background Imager(CBI) \cite{3}, Balloon Observations of Millimetric Extragalactic Radiation and Geophysics (BOOMERanG) \cite{4} and the Planck collaborations \cite{5}. Additionally, ground-breaking progress in cosmology was driven by the observational discoveries of Perlmutter and Riess's teams, shedding new light on the Universe's nature \cite{6,7}. These investigations aim to confirm that the Universe is undergoing accelerated expansion. However, this raises questions about the isotropy of the expansion. Recent findings suggest that the Universe may expand at varying rates in different directions, challenging the notion of isotropic expansion \cite{8,9}. Although FLRW cosmology is highly successful, it relies on the cosmological principle. However, observational evidence suggests slight variations in microwave radiation from different directions, hinting at anisotropies. Bianchi cosmology, which describes anisotropic yet homogeneous geometries, offers a suitable alternative. Various studies have explored Bianchi cosmology within different modified gravity frameworks \cite{10,11,12,13,14}. Holographic dark energy theories present a fascinating interface to probe late-time cosmology, guided by contemporary quantum gravity ideas. This model extends the conventional holographic dark energy framework by incorporating features from fractional calculus recently applied, e.g., in cosmological settings. In this manner, we retrieve a novel form of holographic dark energy density\\
Building on the discovery of black hole thermodynamics \cite{15,16}, Hooft introduced the holographic principle (HP) \cite{17}, which suggests that information within a volume of space can be represented as a hologram on its boundary. This concept, akin to Plato's cave allegory, implies that the information contained in a region can be encoded and understood through its surface representation. The Holographic Principle is now recognized as a foundational concept in quantum gravity and string theory, offering insights into the nature of space-time and gravity. Using the fundamental concepts of quantum gravity, we can explore the nature of dark energy (DE) through the Holographic Dark Energy (HDE) principle. According to the HP, the degrees of freedom in a system are finite and proportional to its surface area, rather than its volume. As a key concept in quantum gravity, the HP has the potential to address long-standing problems in various fields, including the DE problem \cite{18}. The Holographic Principle suggests that physical quantities, such as the energy density of dark energy $(\rho_{d})$, can be represented by values on the universe's boundary. This implies that only two fundamental quantities – the reduced Planck mass ($M_{p}$) and the cosmological length scale ($L$) – are needed to formulate an expression for $\rho_{d}$. This study focuses on the theoretical investigation and observational validation of the LRS Bianchi type-I cosmological model with fractional holographic dark energy. This work examines the relationship between short-distance cutoffs from quantum field theory and long-distance cutoffs imposed by black hole formation constraints. Notably, Cohen et al. \cite{19} demonstrated that holographic considerations yield a specific inequality.
\begin{equation}
\Lambda^{3}L^{3} \leq S^{\frac{3}{4}},
\end{equation}
where $S$ represents the system's entropy, $L$ is the infrared (IR) cutoff, and $\Lambda$ is the ultraviolet (UV) cutoff, with the Planck mass $(m_{p})$ set to 1 for unit simplicity.\\
Recently, several studies have investigated holographic dark energy (HDE) from different perspectives. Alternative forms of HDE have emerged, such as Tsallis HDE models, which modify the standard Boltzmann-Gibbs entropy using Tsallis' corrections, particularly in black hole physics \cite{20,21,22}. In our work, we introduce a new dark energy model that leverages fractional calculus, providing a unique approach to understand dark energy's behaviour. Section II introduces the concept of fractional holographic dark energy, while section III explores the anisotropic and homogeneous tetrad field equations. Section IV presents an analysis of the observational data used to constrain the key parameters of the model. In section V, we analyse the changes in physical quantities. At last, section VI concludes the study by summarising the key findings and presenting the overall conclusions.

 \section{Insight about Fractional Holographic Dark Energy}
 
 To explore the implications of fractional calculus in classical and quantum cosmology, we'll provide a brief overview of its key features. Fractional calculus extends classical differentiation and integration to non-integer orders, making it a valuable tool for studying complex systems. It generalizes derivatives to rational, real, or complex numbers, with various types like Liouville, Riemann, Caputo, and Riesz derivatives \cite{23}. Each type has its own applications, and the choice depends on the specific problem. For more details, refer to the works of Ortigueira \cite{24,25,26}. Fractional calculus has intriguing applications in quantum mechanics, particularly in describing Brownian motion's self-similar, non-differentiable trajectories with fractal dimensions. Feynman and Hibbs leveraged this concept to reformulate quantum mechanics as a path integral over Brownian paths, highlighting the role of fractality \cite{27}. Recent developments led to the creation of a fractional path integral, introducing space-fractional quantum mechanics (FQM). This approach involves path integrals over Lévy flight paths, characterized by the Lévy index $\alpha$. When $\ alpha=2$, it reduces to the Gaussian process or Brownian motion. A key outcome of FQM is the space-fractional Schr$\ddot{o}$dinger equation, where the standard second-order spatial derivative is replaced by a fractional-order derivative, specifically the quantum Riesz fractional derivative. Fractional calculus can be applied to cosmology and gravity, extending existing frameworks to non-integer derivative orders. This approach, known as fractional cosmology, shows promise in addressing challenges like the Hubble tension and cosmological constant problem. It provides a new way to study phenomena such as Universe evolution, black hole dynamics, and gravitational waves \cite{28,29}.\\
 Classical cosmological models using fractional derivatives typically follow two approaches: First-step modification, where a fractional derivative geometry is established first, and Last-step modification, where fractional derivatives replace ordinary ones in field equations. The formal is considered more fundamental. In quantum cosmology, research has focused on deriving the fractional Wheeler-DeWitt equation, inspired by the space-fractional Schr$\ddot{a}$dinger equation, using the quantum Riesz fractional derivative. Jalalzadeh et al.\cite{30} recently explored the impact of fractional quantum mechanics on Schwarzschild black hole thermodynamics. They used a space-fractional Riesz derivative to modify the Wheeler-DeWitt equation and derive fractional black hole entropy. By applying canonical quantization and incorporating the quantum Riesz derivative into the momentum operator, they obtained the fractional Wheeler-DeWitt equation and subsequently calculated the fractional entropy and its mathematical formulation is 
 \begin{equation}
 S_{h} = C A^{\frac{\alpha+2}{2\alpha}}
 \end{equation}
where $\alpha$ represents the fractional parameter whose range is $1 < \alpha \leq 2$ and the constant $C$ is chosen to combine Planck units and reproduce the Bekenstein-Hawking relation when $\alpha = 2$. This equation correspond that the entropy follows a power-law function of its area, similar to Barrow and Tsallis entropies, but with different underlying principles. Combining this with the holographic inequality leads to new insights
\begin{equation}
L^{3}\Lambda^{3} \leq C^{\frac{3}{4}} A^{\frac{3(\alpha +2)}{8 \alpha}}
\end{equation}
Considering $\rho_{d} \sim \Lambda^{4}$, an energy density for holographic dark energy can be proposed, incorporating features of fractional calculus and associated quantum effects as
\begin{equation}
\rho_{d} = \gamma L^{-\frac{(3\alpha -2)}{\alpha}}
\end{equation}
where $\gamma$ is some constant. To align with the original HDE, we put $\gamma = 3 c^{2}$, $c$ is a constant. Hence, the proposed holographic dark energy density can be written as
\begin{equation}
\rho_{d} = 3c^{3}L^{-\frac{3\alpha-2}{\alpha}}
\end{equation}
This new holographic dark energy (HDE) model will be tested for consistency in late-time universal evolution. Although it shares similarities with Barrow and Tsallis models due to power-law entropy-area relations, its motivation differs. Unlike Tsallis' thermodynamic modifications or Barrow's black hole deformations, this model stems from fractional calculus and quantum effects.

\section{ Anisotropic and homogenous  field equations}\label{sec3}
For our cosmological model, we consider the LRS Bianchi type-I metric of the form
\begin{equation}
ds^{2} = -dt^{2} + \mathcal{A}_{1}^{2}(t)dx^{2} + \mathcal{A}_{2}^{2}(t)(dy^{2}+dz^{2})
\end{equation}
where $\mathcal{A}_{1}$ and $\mathcal{A}_{2}$ are the cosmic factors which are function of $t$. \\
Einstein's field equation with a cosmological constant is:
\begin{equation}
R_{\mu \nu} - \frac{1}{2} g_{\mu \nu} R + \Lambda g_{\mu \nu} = -(T_{\mu \nu}+ \overline{T}_{\mu \nu})
\end{equation}
The significance of the notations is as customary. The energy momentum tensor for matter and the new HDE for legitimate interpretation can be put simultaneously as: $T_{\mu \nu} = \rho_{m}u_{\mu}u_{\nu}, \overline{T} = (\rho_{d} + p_{d})u_{\mu}u_{\nu} + g_{\mu \nu}p_{d}$, where $\rho_{m}$ and $\rho_{d}$ represents the energy density with respect to matter and holographic dark energy respectively and $p_{d}$ denotes the HDE pressure. The field equations obtained from $ Eq (7)$ along with the energy momentum tensors by using the metric which is defined in $ Eq (6)$ are
\begin{eqnarray}
2\frac{\ddot{\mathcal{A}_{2}}}{\mathcal{A}_{2}} + \frac{\dot{\mathcal{A}_{2}^{2}}}{\mathcal{A}_{2}^{2}} = -p_{d}\\
\frac{\ddot{\mathcal{A}_{1}}}{\mathcal{A}_{1}} + \frac{\ddot{\mathcal{A}_{2}}}{\mathcal{A}_{2}} + \frac{\dot{\mathcal{A}_{1}}\dot{\mathcal{A}_{2}}}{\mathcal{A}_{1}\mathcal{A}_{2}} = -p_{d}\\
2\frac{\dot{\mathcal{A}_{1}}\dot{\mathcal{A}_{2}}}{\mathcal{A}_{1}\mathcal{A}_{2}} + \frac{\dot{\mathcal{A}_{2}^{2}}}{\mathcal{A}_{2}^{2}} = \rho_{d} + \rho_{m}
\end{eqnarray}
The generalized Hubble parameter $H$ is
\begin{equation}
H = \frac{\dot{a}}{a} = \frac{1}{3}\bigg(\frac{\dot{\mathcal{A}_{1}}}{\mathcal{A}_{1}} + 2\frac{\dot{\mathcal{A}_{2}}}{\mathcal{A}_{2}} \bigg) = \frac{1}{3} (H_{1}+2H_{2})
\end{equation}
we also assume that $\mathcal{A}_{1} \propto \mathcal{A}_{2}^{n}$, where $n$ is a natural number greater than 2. We also derived a relation between the generalized Hubble parameter and the directional Hubble parameter as
\begin{equation}
H_{1} = nH_{2} = \bigg(\frac{3n}{n+2}\bigg)H
\end{equation}
For LRS Bianchi type-I, the field equations from $Eq(8)-Eq(10)$ become after incorporating $ Eq (11)$ and $Eq(12)$
\begin{equation}
\frac{9(2n+1)H^{2}}{(n+2)^{2}} = \rho_{m} + \rho_{d}\\
\end{equation}
\begin{equation}
\frac{6\dot{H}}{(n+2)}+ \frac{27 H^{2}}{(n+2)^{2}} = -p_{d}\\
\end{equation}
\begin{equation}
\frac{9(n^{2}+n+1)H^{2}}{(n+2)^{2}} + \frac{3(n+1)\dot{H}}{(n+2)} = -p_{d}
\end{equation} 
The equation of continuity for dark energy and dark matter are
\begin{eqnarray}
\dot{\rho_{d}} + 3H(1+\omega_{d})\rho_{d} = 0,\\
\dot{\rho_{m}} + 3H(1+\omega_{m})\rho_{m} = 0
\end{eqnarray} 
where $\omega_{d}$ and $\omega_{m}$ are the equation of state parameters (EoS) for dark energy and dark matter, respectively. Here, we do not consider an interacting dark sector. We will proceed under the assumption that dark matter exhibits a pressureless nature, which implies $\omega_{m} = 0$. The selection of the cutoff scale $L$ is a prerequisite for further analysis. The Hubble horizon cutoff, defined by $L = H^{-1}$, was an early proposed approach. This selection sought to mitigate the fine-tuning issue by incorporating a natural length scale linked to the Hubble parameter's inverse $(H^{-1})$. However, it was discovered that this scale caused the dark energy equation of state (EoS) parameter to approach zero and didn't substantially contribute to the Universe's current accelerated expansion. A subsequent proposal suggested adopting the particle horizon as the relevant length scale
\begin{equation}
L_{p} = a \int^{t}_{0} \frac{dt}{a}
\end{equation}
Although this alternative led to an EoS parameter exceeding $-\frac{1}{3}$, it failed to resolve the challenges associated with the Universe's present acceleration. As a result, the future event horizon was explored as a potential length scale
\begin{equation}
L_{f} = a \int^{\infty}_{a} \frac{dt}{a}
\end{equation}
While this scenario can produce the desired acceleration, causality issues arise. A different approach, the Granda-Oliveros cutoff, was introduced, extending the definition of $L$ to include the derivative of $H$ as $L= (\ alpha H^{2} + \beta \dot{H})^{-\frac{1}{2}}$. A more general cutoff framework is provided by the Nojiri-Odintsov cutoff \cite{31}, expressed as $L = (H,\dot{H},\ddot{H},...L_{p}, L_{f},\dot{L}_{p},\dot{L}_{f}...)$. \\
A common concern with these cutoffs is classical stability against perturbations, an issue prevalent across various HDE models with different cutoffs \cite{32}. HDE models can be generalized to include functions of $H$, particle horizon, and event horizon scales. While various forms of HDE, such as Tsallis and Barrow, may approximate a general form, constructing a specific, justified function is typically preferred by Nojiri-Odintsov. In our analysis, we opt for the Hubble horizon cutoff for two reasons. Firstly, it provides a simple framework for fractional HDE (FHDE) to produce universal evolution. Secondly, given the previously demonstrated limitations of the Hubble horizon cutoff, if FHDE can yield observationally consistent results, it would mark significant progress. \\
Applying the Hubble Horizon cutoff $L = H^{-1}$ to $Eq.(4)$ results in
\begin{equation*}
\rho_{d} = \gamma H^{\frac{3\alpha -2}{\alpha}}
\end{equation*}
\begin{equation}
\rho_{d} = 3c^{2} H^{\frac{3\alpha -2}{\alpha}}
\end{equation}
We also define the fractional density parameter for DE and DM as follows
\begin{equation}
\Omega_{d} \hspace{0.25cm}= \hspace{0.25cm} \frac{(n+2)^{2}\rho_{d}}{9(2n+1)H^{2}}\hspace{0.25cm} = \hspace{0.25cm}\frac{c^{2}(n+2)^{2}}{(2n+1)} H^{\frac{\alpha-2}{\alpha}}\\
\end{equation}
\begin{equation}
\Omega_{m} \hspace{0.25cm}= \hspace{0.25cm} \frac{\rho_{m}}{3H^{2}} \frac{(n+2)^{2}}{9(2n+1)}
\end{equation}
From $Eq.(13)$, we conclude that 
\begin{equation}
\Omega_{d} + \Omega_{m} = \Omega_{d}(1+y) = 1
\end{equation}
where $y= \frac{\Omega_{m}}{\Omega_{d}}$ and by using $Eq.(13)-Eq.(15)$, we get
\begin{equation}
\frac{\dot{H}}{H^{2}} = -\frac{3}{2}(1+y+\omega_{d})\Omega_{d}
\end{equation}
By incorporating $Eq.(16)$ and $Eq.(20)$, we get
\begin{equation}
\frac{\dot{H}}{H^{2}} = -\frac{3\alpha}{3\alpha -2} (1+\omega_{d})
\end{equation}
From $Eq.(24)$ and $Eq.(25)$, we get
\begin{equation}
\omega_{d} = -1+  \frac{(3\alpha-2)(1-\Omega_{d})}{2\alpha-(3\alpha-2)\Omega_{d}}
\end{equation}
By using the relation,$d\ln{a} = -\frac{dz}{(1+z)}$ in  $Eq.(16)$, and considering $\omega_{d}$ as a constant, we get
\begin{equation}
\rho_{d} = \rho_{d_{0}}(1+z)^{3(1+\omega_{d})}
\end{equation}
And similarly in  $Eq.(17)$, after considering DM has pressureless form, means $\omega_{m} =0$, we get
\begin{equation}
\rho_{m} = \rho_{m_{0}}(1+z)^{3}
\end{equation}
After incorporating $Eq.(27)$ and $Eq.(28)$ in $Eq.(13)$, we get parameterized form of Hubble parameter is
\begin{equation}
H(z) = H_{0}[\Omega_{m_{0}}(1+z)^{3} + (1-\Omega_{m_{0}})(1+z)^{3(1+\omega_{d})}]
\end{equation}
To further investigate the cosmological parameter in Eq. (29), the optimal values of $H_{0},\alpha$ and $\Omega_{m_{0}}$ will be determined using different datasets combinations: Hubble, Hubble + BAO, Hubble + Pantheon+ Shoes, and Hubble + BAO + Pantheon+Shoes datasets. From $Eq.(26)$, we take the value of $\omega_{d}$.

\section{Constraints on observational methodologies and validation of findings}

Precise Universe modeling relies on high-quality observational data and reliable parameter estimation techniques. To validate our model's accuracy, we utilize modern datasets to validate the model's parameter accuracy, leveraging 46 Hubble measurements, 15 Baryon Acoustic Oscillation (BAO) observations, and supernova distances from the extensive Pantheon+Shoes datasets of 1701 entries to constrain optimal parameter values. By integrating these varied datasets, we constrain model parameters, enabling a detailed investigation into the Universe's evolution. Furthermore, we examine the LRS Bianchi type-I cosmological model, incorporating holographic dark energy to gain deeper insights. We employ the Scipy optimization method in Python to analyse the datasets and refine parameter values using the emcee Python framework for Markov Chain Monte Carlo (MCMC) analysis.\\

\subsection{$H(z)$ data-sets}

Analysing the Hubble parameter is crucial for understanding the expanding Universe, and its relation to the redshift parameter $z$ proves particularly valuable in various contexts. By examining specific redshift values, we can deduce the Hubble parameter's value. The Hubble parameter as a function of redshift, $H(z)$, can be derived from the redshift-cosmological time relation, expressed as $H(z) = -\frac{1}{(1+z)}\frac{dz}{dt}$. This formulation enables the determination of $H(z)$ through spectroscopic surveys that precisely measure the differential time $dt$ corresponding to a redshift interval $dz$. The empirical determination of $H(z)$ at various redshift typically employs two main methodologies.\\
\begin{enumerate}
                 \item  $\textbf{Baryon Acoustic Oscillation(BAO)}$ This approach utilises the BAO scale, a feature imprinted on galaxy distribution by early Universe sound waves, serving as a standard ruler. By measuring this scale across different redshifts, the Hubble parameter can be inferred \cite{33}.
                 \item  $\textbf{Differential age method}$ This method leverages the age differences among passively evolving galaxies at distinct redshifts to constrain the Hubble parameter. By measuring the redshift difference $dz$ from galaxy spectra and the corresponding differential age $dt$ via stellar population  method, the technique utilises the relation $H(z) = -\frac{1}{(1+z)}\frac{dz}{dt}$.\\   
               \end{enumerate}
 
Our analysis utilise 46 Hubble parameter measurements over the redshift range $0 \leq z \leq 2.36$ to constrain a cosmological model. We employ maximum likelihood estimation equivalent to chi-square minimisation to determine the values of key parameters $H_{0}, \alpha$, and $\Omega_{m_{0}}$. The chi-square statistic is defined as:
\begin{equation}
\chi^{2}_{H}(H_{0}, \alpha,\Omega_{m_{0}} ) = \sum^{46}_{i=1} \frac{[H_{th}(z_{i},H_{0}, \alpha,\Omega_{m_{0}})-H_{ob}(z_{i})]^{2}}{\sigma^{2}_{H}(z_{i})}
\end{equation} 
In this context, it $H_{th}$ represents the theoretically predicted Hubble parameter, $H_{ob}$ is the observed value, and $\sigma$ denotes the standard deviation of the measurement. $Fig.1$ displays the corresponding contour plots for model parameters, indicating confidence regions at $1-\sigma$ and $2-\sigma$ levels, and $Fig.2$ presents the distribution of observational Hubble data points with their associated errors. By analysing these datasets, we can constrain cosmological parameters and evaluate the proposed model's effectiveness in describing the Universe's expansion history.\\

\begin{figure}[h!]
\includegraphics[height=3in]{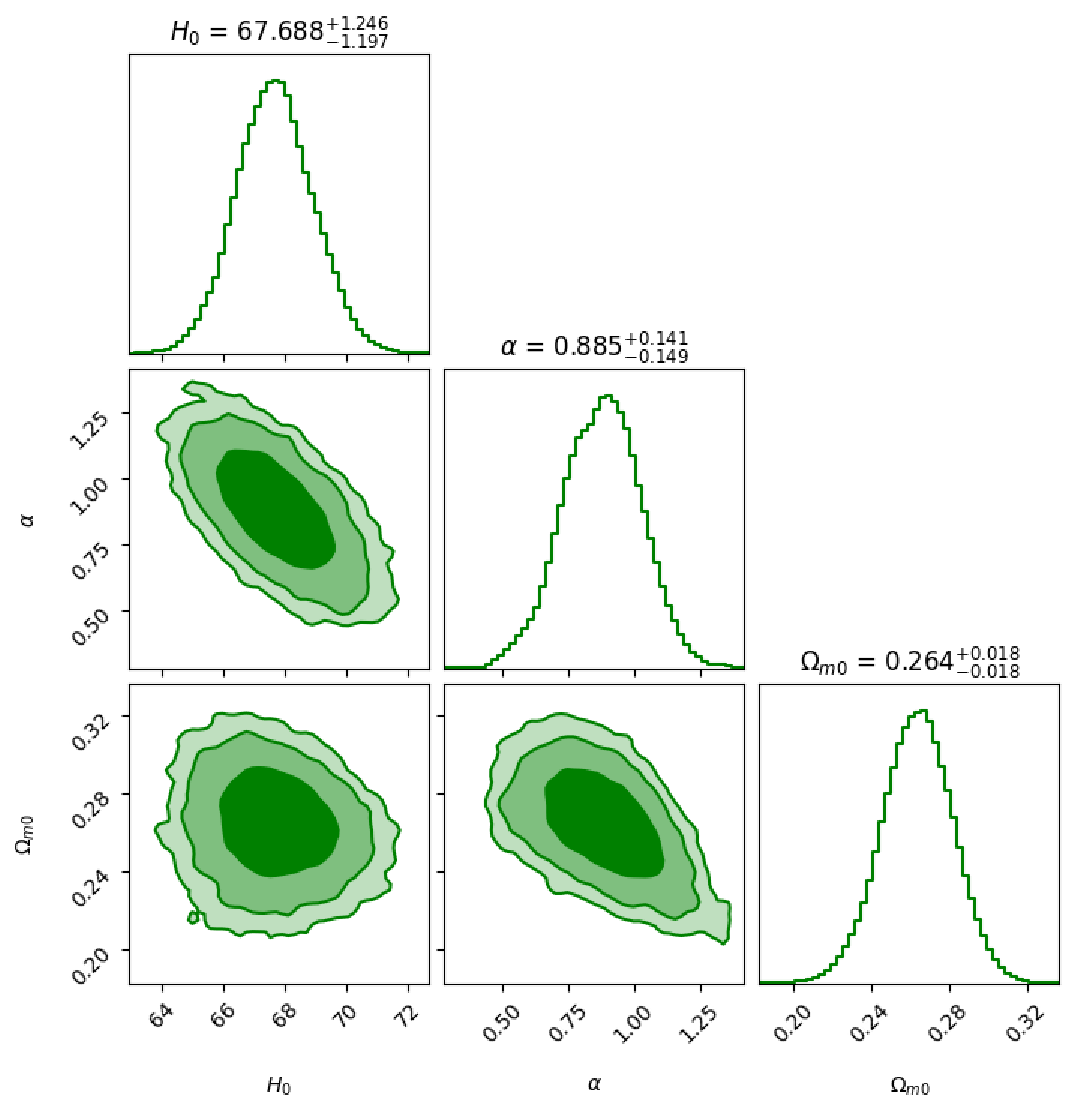}

{{Figure 1:}  Contour plot with $1-\sigma$ and $2-\sigma$ confidence regions for the parameter $H_{0}, \alpha$ and $\Omega_{m0}$ along with the constraint values for Hubble datasets.}
\end{figure}

\begin{figure}[h!]
\includegraphics[height=3in]{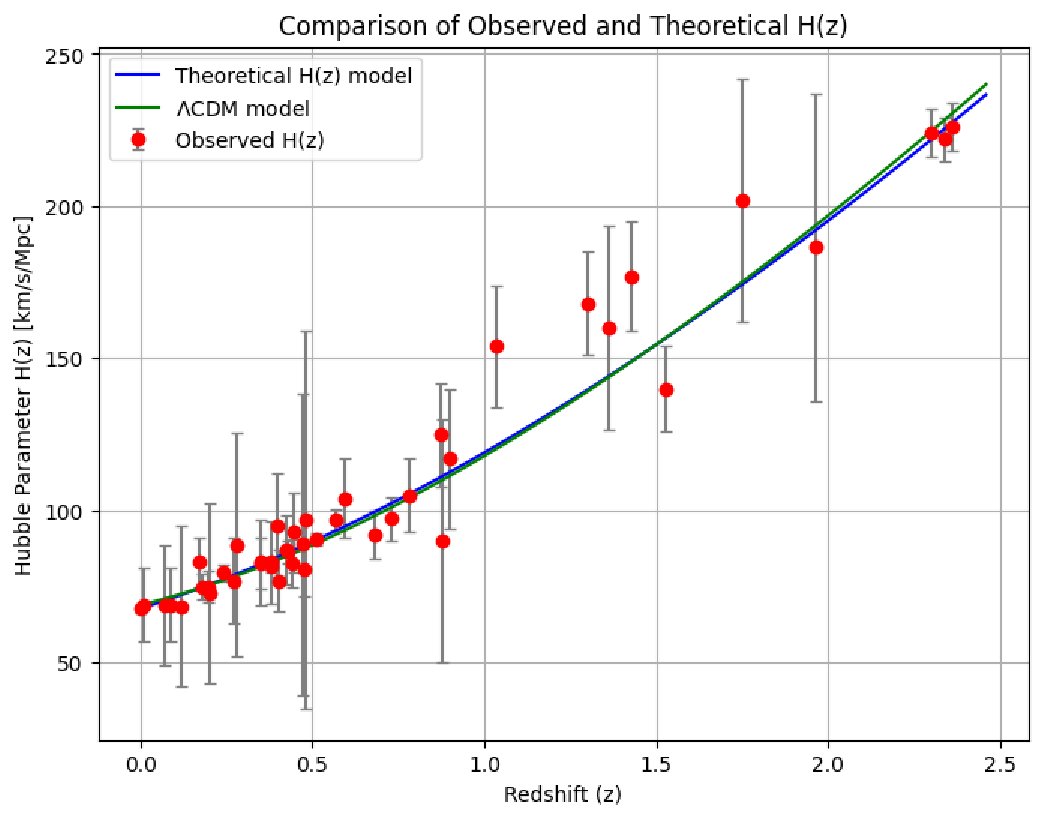}

{{Figure 2:}  $H(z)$ is shown as a function of redshift $z$, with 46 observational data points and their corresponding error bars, for the chosen cosmological model}
\end{figure}
\subsection{BAO datasets}

We studied the oscillations induced by cosmic perturbations in the early Universe by analysing a fluid consisting of photons, baryons, and dark matter, coupled through Thompson scattering. To measure BAO, we employed three surveys: The Six Degree Field Galaxy Survey (6dFGS), the High-resolution Sloan Digital Sky Survey (SDSS), and the Baryon Oscillation Spectroscopic Survey (BOSS). From 330 BAO data points, we selected 15 key representative points to minimise errors arising from dataset correlations.\\
We leverage transverse BAO experiments and the comoving angular diameter distance to probe the Universe's expansion history and constrain cosmological models. The mathematical formulas used to analyse the BAO datasets are as follows:
\begin{equation}
d_{A}(z) = c \int^{z}_{0} \frac{dz^{'}}{H(z^{'})}
\end{equation}  
\begin{equation}
D_{v}(z) = \bigg[\frac{d_{A}(z)^{2}cz}{H(z)}\bigg]^{\frac{1}{3}}
\end{equation}
where $d_{A}(z)$ represents the comoving angular diameter distance, $D_{v}(z)$ denotes the dilation scale, and $c$ stands for the covariance matrix \cite{34}. For the BAO dataset, the $\chi^{2}$ function is defined as
\begin{equation}
\chi^{2}_{BAO} = \sum_{i=1}^{15}\bigg[\frac{D_{th}(z_{i})-D_{obs}}{\Delta D_{i}}\bigg]^{2}
\end{equation}
Within the redshift interval $0.24 < z < 2.36$. The equation evaluates the model's theoretical distance modulus $D_{th}$ against observed values $D_{obs}$ at various redshifts, $z_{i}$ where it $i$ ranges from 1 to $n$, leveraging data from BAO observations.

\subsection{Integrated data analysis of Hubble and BAO datasets}
We employ a joint analysis of Hubble and BAO datasets via the MCMC method to improve parameter estimation precision in the LRS Bianchi type-I cosmological model with fractional holographic dark energy. Combining these datasets leverages their complementary strengths: Hubble data directly probes the Universe's expansion history, while BAO data constrains cosmic structure scales. This approach enhances model parameter accuracy and robustness through chi-square minimization. The chi-square function is defined as
\begin{equation}
\chi^{2} = \chi^{2}_{H} + \chi^{2}_{BAO}
\end{equation}
where $\chi^{2}_{H}$ and $\chi^{2}_{BAO}$ are the chi-square distributions with respect to Hubble datasets and BAO datasets, respectively. This integrated framework enables us to determine the most likely model parameter values by combining the complementary constraints from both observational datasets. The joint analysis yields contour plots $Fig.3$ showing $1-\sigma$ and $2-\sigma$ confidence regions for the model parameters, representing the confidence level of the estimates. Additionally, $Fig.4$'s error bar plot demonstrates the fit of our $H(z)$ model to the combined datasets, with 46 Hubble data points and 15 BAO data points.

\begin{figure}
\includegraphics[height=3in]{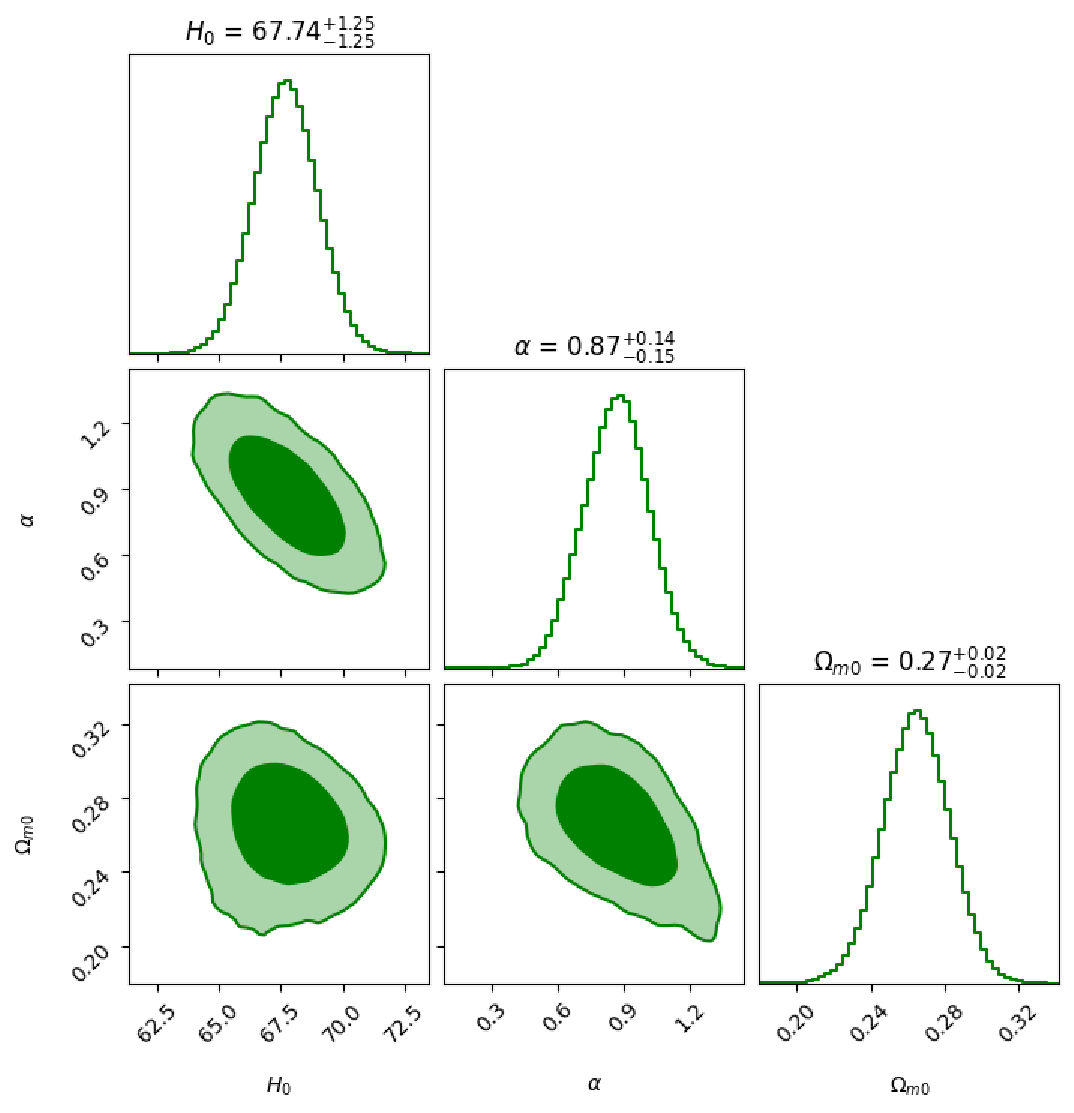}

{{Figure 3:}  Contour plot with $1-\sigma$ and $2-\sigma$ confidence regions for the parameter $H_{0}, \alpha$ and $\Omega_{m0}$ along with the constraint values for Hubble+BAO datasets.}
\end{figure}
\begin{figure}[h!]
\includegraphics[height=3in]{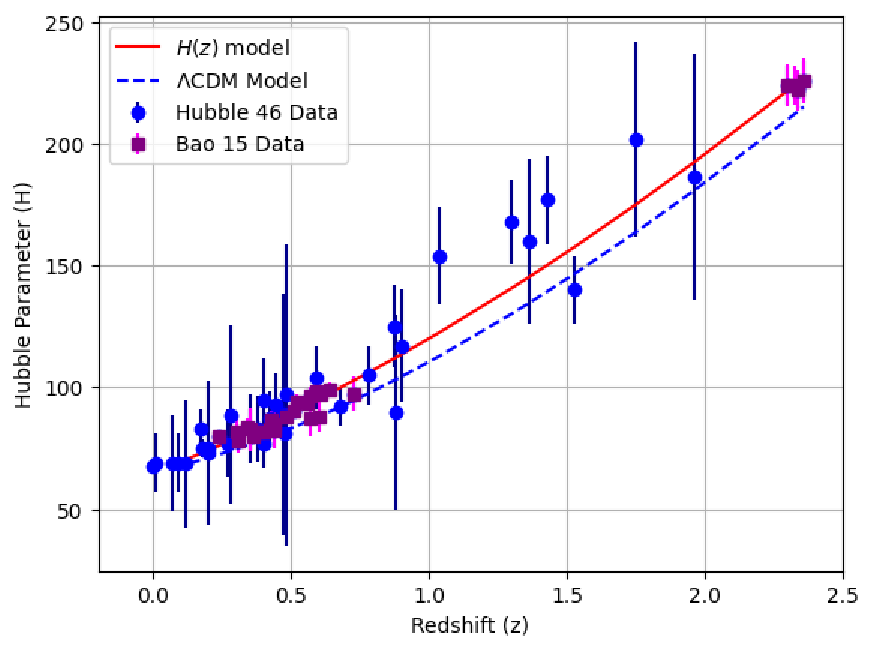}

{{Figure 4:}  $H(z)$ +BAO vs. error bars plot, for the chosen cosmological model}
\end{figure}

\subsection{Pantheon + Shoes datasets}

The Pantheon +Shoes datasets, consisting of 1701 light curve of 1550 distinct type of supernovae data points spanning redshifts from 0.02012 to 2.2613, offers a robust framework for investigating the Universe's expansion history. The datasets provide precise measurements of luminosity and comoving distances, critical for testing cosmological models and understanding cosmic acceleration's underlying dynamics \cite{35,36}. We utilise the extensive and accurate data from the Pantheon datasets to refine our LRS Bianchi type-I cosmological model in the presence of fractional holographic dark energy, thereby imposing stringent constraints on the model's parameters.\\
The comoving distance $D_{M}(z)$ is a fundamental concept in cosmology, representing the proper distance between two points in the Universe at a given time, accounting for the expansion of space. It is defined as
\begin{equation}
D_{M}(z) = \int^{z}_{0} \frac{c dz^{'}}{H(z^{'})}
\end{equation}
where $c$ the speed of light and $H(z)$ is the Hubble parameter at redshift $z$. This definition allows for accurate calculations of distances in an expanding Universe. The comoving distance provides valuable insights into the Universe's expansion history, shedding light on the behaviour of components like dark energy and matter. Another key metric is the luminosity distance $D_{L}(z)$, which relates to the observed brightness of distant objects like supernovae. The luminosity distance is defined as:
\begin{equation}
D_{L}(z) = (1+z) \int^{z}_{0} \frac{c}{H(z^{'})} dz^{'} \hspace{0.75cm} \approx  \hspace{0.5cm}  D_{L}(z) = (1+z)D_{M}(z)
\end{equation} 
where $D_{M}(z)$ the transverse comoving distance. This relationship enables the use of standard candles, such as supernova, to probe the expansion history and geometry of the Universe. The relationship between observed distance modulus $\mu$ and the luminosity distance $D_{L}(z)$ is given by
\begin{equation}
\mu = 5\log_{10}\bigg(\frac{D_{L}}{H_{0}Mpc}\bigg) + 25
\end{equation}
This equation shows that the distance modulus $\mu$ is directly related to the luminosity distances $D_{L}(z)$, enabling the use of standard candles like supernovae to measure cosmic distances.\\
For Pantheon datasets, the chi-square is expressed as 
\begin{equation}
\chi^{2}= \sum_{k=1}^{1701}\bigg[\frac{\mu_{th}(z_{k})-\mu_{obs}(z_{k})}{\sigma_{\mu}(z_{k})}\bigg]^{2}
\end{equation}
where $\mu_{obs}(z_{k})$ is the observed distance modulus at redshift $z_{k}$, $\mu_{th}(z_{k})$ is the theoretical distance modulus and $\sigma_{\mu}(z_{k})$ denotes the uncertainty in the observed distance modulus.

\subsection{Hubble and Pantheon+Shoes combine datasets analysis}
To refine parameter estimation and deepen understanding of the LRS Bianchi type-I cosmological model with fractional holographic dark energy, we conduct a joint analysis combining the Hubble and Pantheon + Shoes datasets. This approach harnesses the complementary strengths of each dataset, yielding more robust and reliable model parameter constraints. The joint analysis minimizes the overall chi-square function by combining contributions from both datasets. And the chi-square function is
\begin{equation}
\chi^{2} = \chi^{2}_{H} + \chi^{2}_{SNe}
\end{equation}
here, $\chi^{2}_{H}$ and so $\chi^{2}_{SNe}$ are the chi-square distributions with respect to Hubble datasets and Pantheon+ Shoes datasets consecutively. The joint analysis yields tighter constraints on the model parameters $H_{0}, \alpha$, and $\Omega_{m_{0}}$, resulting in a more precise and well-defined parameter space. By combining the strengths of both datasets, we reduce uncertainties and enhance model reliability. The contour plot $Fig.5$ illustrates the $1-\sigma$ and $2-\sigma$ confidence regions for the model parameters based on the combined Hubble and Pantheon datasets.
\begin{figure}
\includegraphics[height=5in]{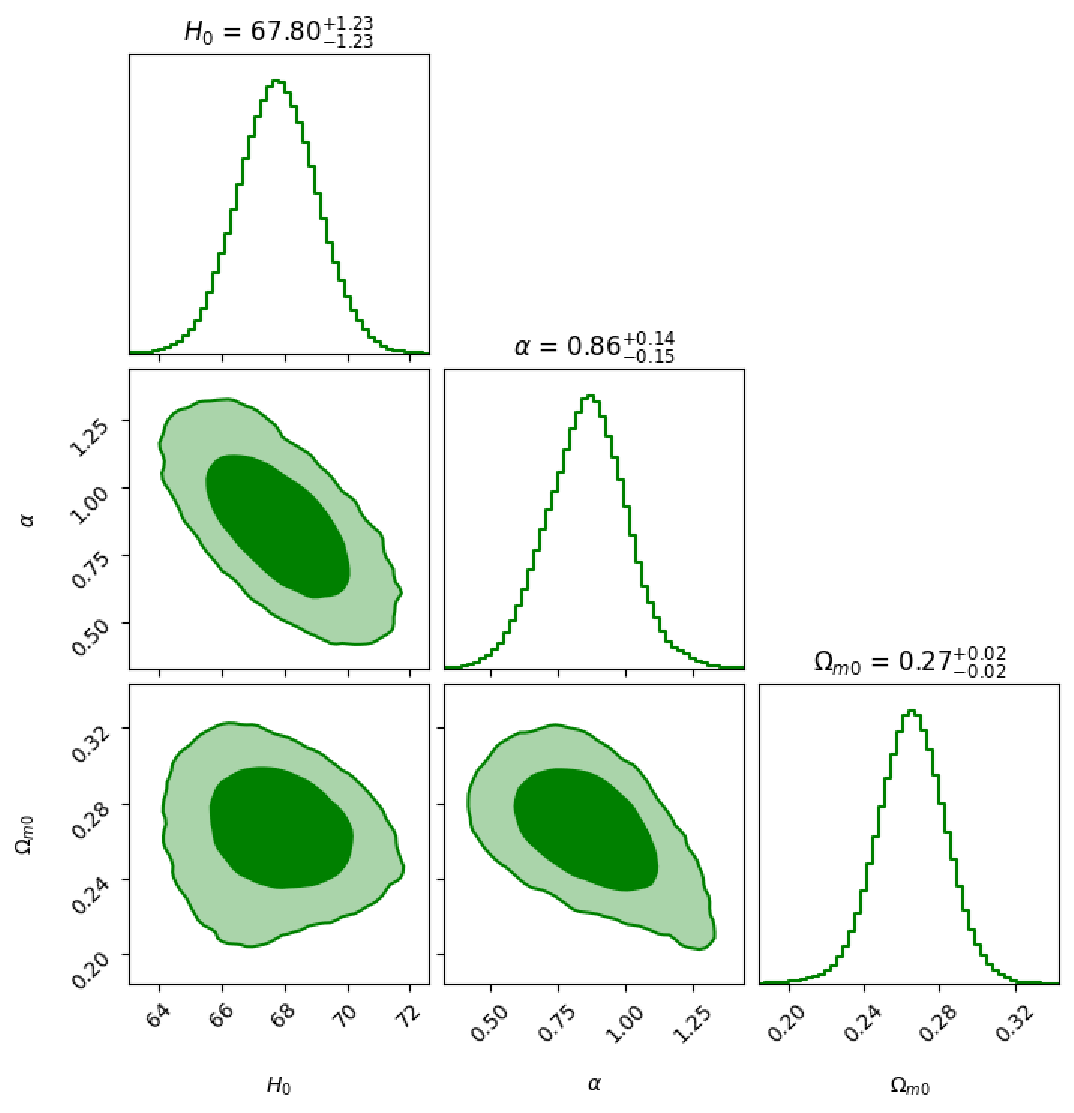}

{{Figure 5:}  Contour plot with $1-\sigma$ and $2-\sigma$ confidence regions for the parameter $H_{0}, \alpha$ and $\Omega_{m0}$ along with the constraint values for Hubble+Pantheon datasets.}
\end{figure}

\subsection{Comprehensive analysis by integrating data from Hubble, BAO and Pantheon+ Shoes datasets }

To refine parameter estimation and deepen our understanding of the LRS Bianchi type-I cosmological model in the presence of fractional Holographic dark energy, we conduct a joint analysis combining Hubble, BAO, and Pantheon+Shoes (SNe) datasets. This approach capitalises on the unique strength of each dataset, yielding complementary constraints that enhance the robustness and reliability of our parameter determination. The joint analysis involves merging the chi-square contributions from each datasets, enabling us to minimize the total chi-square function.
\begin{equation}
\chi^{2} = \chi^{2}_{H} + \chi^{2}_{BAO} + \chi^{2}_{SNe}
\end{equation}
where $\chi^{2}_{H}$ denotes the chi-square distribution from the Hubble datasets, $\chi^{2}_{BAO} $ denotes the chi-square distribution from BAO datasets and $\chi^{2}_{SNe}$ is the chi-square distribution with respect to Pantheon+ Shoes datasets. The joint analysis yields more precise constraints on the model parameter $H_{0},\alpha$, and $\Omega_{m_{0}}$. By synergistically combining the datasets, we obtain a better-defined parameter space, characterised by reduced uncertainties and enhanced model reliability. $Fig.6$ showcase the contour plot indicating the $1-\sigma$ and $2-\sigma$ regions for the integrated Hubble, BAO and Pantheon+Shoes datasets.\\
We carefully selected the Hubble 46, BAO15, and 1701 Pantheon + Shoes datasets to comprehensively examine the cosmic expansion. The Hubble datasets provide direct $H(z)$ measurements, tracing the expansion history across various redshifts. BAO measurements serve as a standard ruler for large scale structures, constraining expansion rates and distance scales. The Pantheon+Shoes datasets probe late-time acceleration, shedding light on dark energy's behaviour. Combining Hubble and BAO data yields robust constraints on the Universe's early and intermediate expansion, validating findings on transition redshifts and deceleration parameters. In contrast, the Hubble-Pantheon+Shoes dataset emphasises late-time cosmic behaviour, shedding light on dark energy's equation of state and its influence on accelerated expansion.
\begin{table}
\centering
\caption{ Estimated model parameter $H_{0},\alpha$ and $\Omega_{m_{0}}$}
\begin{tabular}{|p{3cm}|p{2.5cm}|p{2.5cm}|p{2.5cm}|}
\hline
Parameters & $H_{0}$  & $\alpha$ & $\Omega_{m_{0}}$   \\
\hline
$H(z)$ & $67.688^{+1.246}_{-1.197}$ & $0.885^{+0.141}_{-0.149}$ & $0.264^{+0.018}_{-0.018}$  \\
\hline
$H(z)$+BAO & $67.74^{+1.25}_{-1.25}$ & $0.87^{+0.14}_{-0.15}$ & $0.27^{+0.02}_{-0.02}$ \\
\hline
$H(z)$+SNe & $67.80^{+1.23}_{-1.23}$ & $0.86^{+0.14}_{-0.15}$ & $0.27^{+0.02}_{-0.02}$   \\
\hline
$H(z)$+BAO+SNe & $67.73^{+1.23}_{-1.24}$ & $0.87^{+0.14}_{-0.14}$ & $0.27^{+0.02}_{-0.02}$\\ 
\hline

\end{tabular}\\
\end{table}
The combination of these datasets allows for a comprehensive exploration of the Universe's evolution, leveraging their complementary strengths and reliability. By integrating multiple datasets, we decrease the impact of individual datasets biases and enhance the statistical significance of our model parameters. This multi-dataset methodology strengthens the reliability of our conclusions and minimises parameter degeneracy, aligning with the contemporary research in modified gravity theories that employ combined datasets to explore dark energy and cosmic acceleration. By adopting this approach we establish a more extensive and reliable basis for testing LRS Bianchi type-I cosmological models, generating results that are both robust and comparable to existing studies \cite{37,38,39}. We employ the MCMC to estimate the value of $H_{0}, \alpha$ and $\Omega_{m_{0}}$, with the outcomes detailed in $Table 1$. Analysis of the Hubble datasets resulted in the following constrained values for the parameters: $H_{0} = 67.688^{+1.246}_{-1.197}, \alpha = 0.885^{+0.121}_{-0.149}$ and $\Omega_{m_{0}} = 0.264_{-0.018}^{+0.018}$. From Hubble + BAO datasets, we get the value of parameters as $H_{0} = 67.74^{+1.25}_{-1.25} , \alpha = 0.87^{+0.14}_{-0.15}$ and $\Omega_{m_{0}} = 0.27_{-0.02}^{+0.02}$. From the integrating datasets of Hubble and Pantheon+Shoes we get, $H_{0} = 67.80^{+1.23}_{-1.23} , \alpha = 0.86^{+0.14}_{-0.15}$ and $\Omega_{m_{0}} = 0.27_{-0.02}^{+0.02}$ which is inline with Planck data. At last we do the combine analysis of the Hubble, BAO and Pantheon+Shoes datasets; we get the values of our model parameters $H_{0}, \alpha$ and $\Omega_{m_{0}}$ as $67.73^{+1.23}_{-1.24}, 0.87^{+0.14}_{-0.14} $ and $0.27_{-0.02}^{+0.02}$ respectively. Our results are in line with the well-established observational evidence \cite{40}. Our results' alignment with existing references validates our analysis and confirms the proposed model's reliability within current observational data. By yielding $H_{0}$ values consistent with Planck's findings, our research bolsters the LRS Bianchi type-I model's credibility, particularly when incorporating fractional holographic dark energy, as a plausible explanation for cosmic acceleration.
\begin{figure}     
\includegraphics[height=6in]{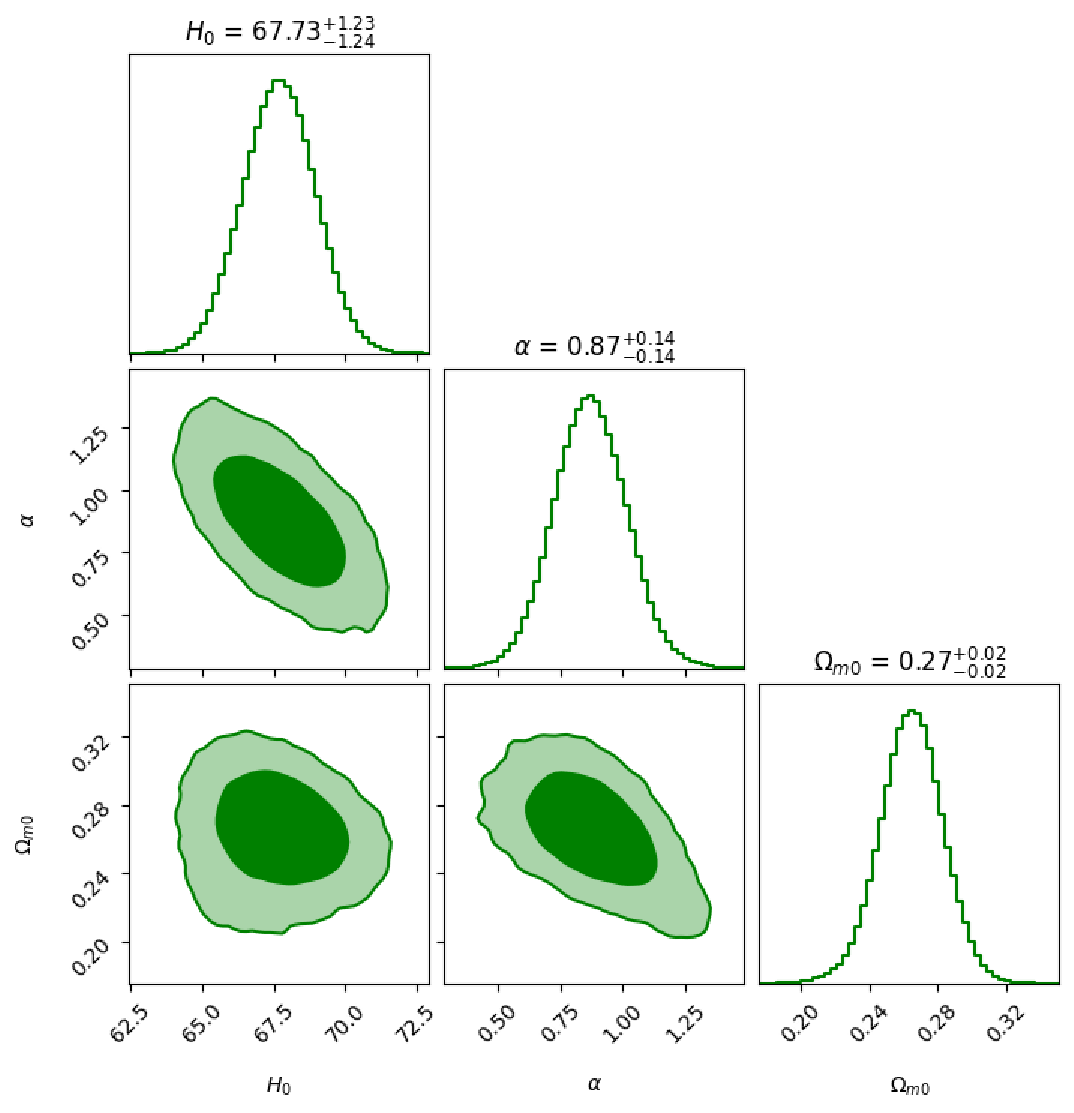}

{{Figure 6:}  Contour plot with $1-\sigma$ and $2-\sigma$ confidence regions for the parameter $H_{0}, \alpha$ and $\Omega_{m0}$ along with the constraint values for Hubble+BAO+Pantheon datasets.}

\end{figure}
\section{Analysis of the Changes in Physical Quantities}
This section describes the trends and behavior of the density parameter with dark energy and dark matter, the equation of state (EoS) parameter, deceleration parameter, and state-finder parameter.
\subsection{Density parameter}
The density parameter is a dimensionless quantity that characterises the contribution of different energy components-such as matter, radiation, and dark energy—to the total energy density of the universe. The density parameter associated with fractional holographic dark energy is denoted by $\Omega_{d}$. The value of $\Omega_{d}$ plays a crucial role in determining the dynamics and fate of the universe. Observational evidence, such as data from Hubble, Hubble+BAO, Hubble+Pantheon+Shoes, and Hubble+BAO + SNe strongly supports that the present value of $\Omega_{d}$ is approximately $0.73$, indicating that dark energy constitutes about $ 70\%$  of the total energy content of the universe. By using $Eq.(21)$, we can calculate the value of $\Omega_{d}$ in terms of redshift $z$
\begin{equation}
\bigg(\frac{\Omega_{d}}{\Omega_{d_{0}}}\bigg)^{2\alpha}\bigg(\frac{1-\Omega_{d_{0}}}{1-\Omega_{d}}\bigg)^{(2-\alpha)} = (z+1)^{3(\alpha-2)}
\end{equation} 
where $\Omega_{d_{0}}$ denotes the $\Omega_{d}$ when $z=0$. And we take the value of $\Omega_{d_{0}} = 0.73$ from the observational analysis which we perform in the previous sections.
 In $Fig.(7)$, we have plotted the fractional holographic dark energy density parameter $\Omega_{d}$ as a function of redshift $z$, using the value $\alpha = 1.01$ which is based on observational analysis. The plot shows that a reasonable evolution of the dark energy density can be obtained, especially as fractional characteristics' effects become more significant later.\\
\begin{figure}     
\includegraphics[height=3in]{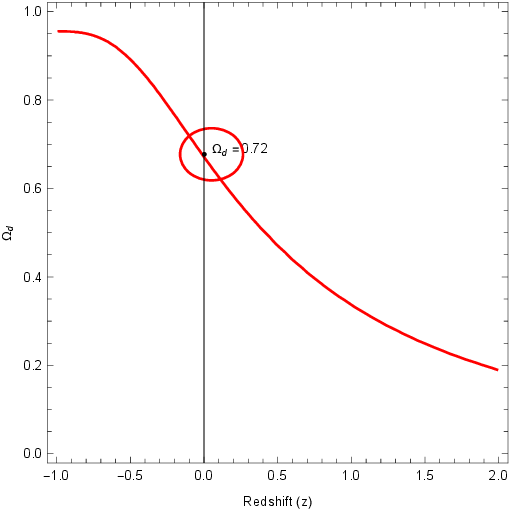}

{{Figure 7:}  The change in the dark energy density parameter $\Omega_{d}$ with redshift $z$ is shown for the value $\alpha=1.01$}

\end{figure}
Similarly, we plot for the density parameter associated with dark matter, which is denoted by $\Omega_{m}$. By using $Eq.(23)$, we get an expression for $\Omega_{m}$. It represents the contribution of matter to the total energy density of the universe, normalized by the critical density. This dimensionless parameter tells us what fraction of the total energy density is made up of matter. 
\begin{figure}     
\includegraphics[height=3in]{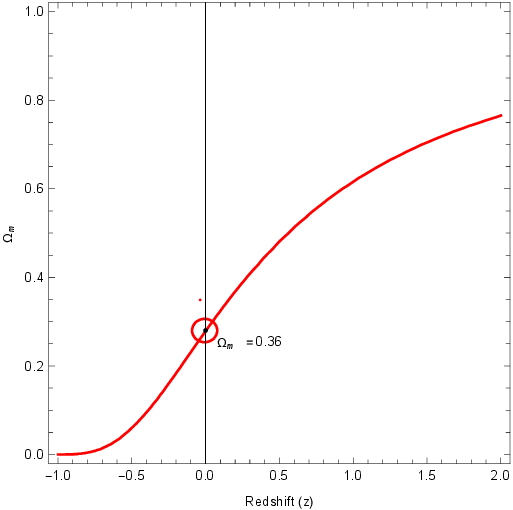}

{{Figure 8:}  The change in the dark matter density parameter $\Omega_{m}$ with redshift $z$ is shown for the value $\alpha=1.01$}

\end{figure}
\subsection{Deceleration parameter}
The deceleration parameter $q$ is an important quantity in cosmology that helps describe the nature of the Universe’s expansion. It indicates whether the expansion is speeding up or slowing down. This parameter is defined using the Hubble parameter $H$ and its time derivative $\dot{H}$. Mathematically, it is written as: 
\begin{equation}
q = -1-\frac{\dot{H}}{H^{2}}
\end{equation}
Here, $\dot{H}$ is the rate at which the Hubble parameter changes with time. By using $Eq.(25)$ and $Eq.(26)$, the expression for $q$ is 
\begin{equation}
q = -1 + \frac{3\alpha (1-\Omega_{d})}{2\alpha - \Omega_{d}(3\alpha - 2)}
\end{equation} 
where $\alpha$ and $\Omega_{d}$ are the model parameters and their values can be obtained from the observational analysis, which we performed in the previous sections. Based on the previous analysis, we get the value of $\alpha$ and $\Omega_{d}$ are 1.01 and 0.73 respectively which are in line with the observational evidence \cite{41}. 
 \begin{figure}[h!]    
\includegraphics[scale=1]{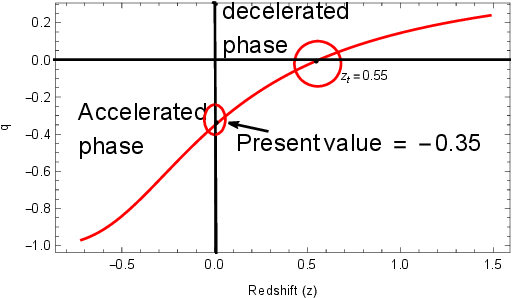}

{{Figure 9:}  The change in the deceleration parameter $q$ with redshift $z$ is shown for the value $\alpha=1.01$}

\end{figure}
We plotted a graph for $q$ in $Fig(9)$ after using the value of the model parameters, which we obtained from observational analysis. From $Fig.9$, the deceleration parameter $q$ begins with a positive value, indicating a decelerating phase in the early Universe. At the transition redshift $z_{t} = 0.55$, the Universe enters a transition phase from deceleration to acceleration. This value lies within the observed range $0.5 \leq z_{t} \leq 0.8$, consistent with current observational data. As time progresses,  $q$ gradually approaches $-1$, representing the present accelerated expansion of the Universe.

\subsection{Equation of State parameter}
The equation of state (EoS) parameter is an important concept in cosmology that describes the relationship between pressure and energy density of various components in the Universe. It plays a key role in understanding how the Universe expands and why this expansion is accelerating. For example, $\omega_{d}=0$ corresponds to dust or non-relativistic matter, $\omega_{d} = \frac{1}{3}$ represents radiation, and $\omega_{d} = -1$ refers to the cosmological constant, as seen in the $\Lambda$CDM model. When $\omega_{d} < -1$, it indicates the phantom phase, while the range $-1 < \omega_{d} < -\frac{1}{3}$ represents the quintessence phase. For our cosmological model, the expression of EoS is given in  $Eq.(26)$. For our model, the value of EoS parameter depends on $\alpha$ and $\Omega_{d}$. From $Fig.(10)$, at early times (i.e., high redshift), the equation of state parameter $\omega_{d}$ begins near zero, indicating a matter-dominated Universe. As time moves forward (towards lower redshift), $\omega_{d}$ shifts to negative values, suggesting a transition to a quintessence-like phase where dark energy starts to dominate. This evolution of $\omega_{d}$ reflects a shift in the Universe's behavior—from a decelerating expansion to an accelerating one. From $Fig.(10)$, it is evident that the EoS parameter evolves from a positive value, indicating a stiff matter-dominated phase in the early Universe. It then passes through the matter-dominated region and gradually approaches $\omega_{d}\rightarrow -1$ at late times. This behavior resembles a de Sitter phase and is consistent with the predictions of the $\Lambda$CDM model.

\begin{figure}[h!]
\centering    
\includegraphics[scale=0.7]{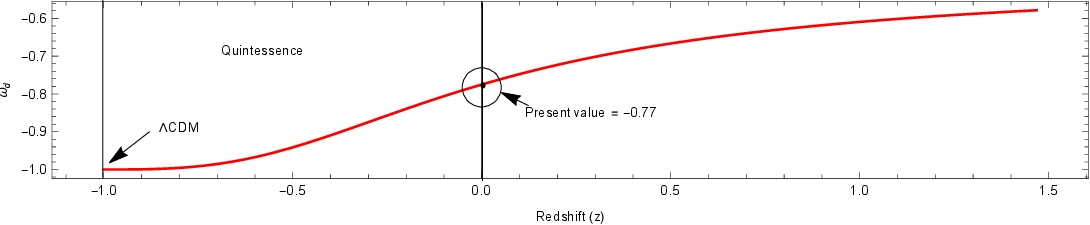}

{{Figure 10:}  Equation of State parameter $\omega$ with redshift $z$ is shown for the value of $\alpha = 1.01$}

\end{figure}

\subsection{State-finder parameter}
The statefinder parameters ${r,s}$ serve as valuable diagnostic tools for comparing the dynamical behaviour of various DE models in a model-independent way. Unlike traditional parameters such as the Hubble parameter $H$ and the deceleration parameter $q$, the statefinder approach involves higher-order derivatives of the cosmic scale factor to detect more subtle differences in cosmic evolution \cite{42}. The parameter $r$ represents the dimensionless third derivative of the scale factor, given by $\frac{\dddot{a}}{aH^{3}}$, and is often referred to as the "jerk" parameter. It provides information about the rate of change of acceleration in the Universe. Different dark energy models predict distinct behaviors for $r$, making it useful for distinguishing between them. The parameter $s$ is defined in terms of both $r$ and $q$, and acts as a measure of how the deceleration parameter evolves. It helps characterize the nature of cosmic acceleration and how it changes over time. Together, the statefinder pair ${(r,s)}$ allows for the classification and comparison of dark energy models without requiring specific assumptions about their underlying physics, making it a powerful tool in modern cosmology.
\begin{figure}     
\includegraphics[height=3in]{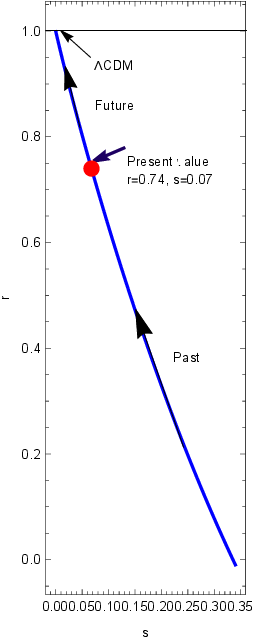}

{{Figure 11:}  State-finder parameter with redshift $z$ is shown for the value of $\alpha = 1.01$}

\end{figure}
The statefinder parameters $(r,s)$ provide a useful way to distinguish between different cosmological models based on how the Universe expands over time. For instance, the $\Lambda$CDM model, characterised by $(r=1,s=0)$, represents a Universe dominated by a cosmological constant. In this scenario, the Universe experiences accelerated expansion, but the rate of acceleration remains constant. In contrast, the quintessence model is described by $(r<1,s>0)$ and reflects a form of dynamical dark energy where the acceleration changes over time, indicating that the dark energy density is not constant. Another example is the Chaplygin Gas model, which has $(r>1,s<0)$. This model involves an exotic type of fluid that unifies both dark matter and dark energy, leading to a unique acceleration behaviour. Finally, the SCDM (Standard Cold Dark Matter) model is identified by $(r=1,s=1)$, which corresponds to a Universe dominated entirely by matter, resulting in a decelerating expansion without any dark energy influence. These classifications help in understanding and comparing the expansion history predicted by various dark energy models. The statefinder parameters in cosmology are defined as
\begin{equation}
r = \frac{\dddot{a}}{aH^{3}} = q+ 2q^{2} - \frac{\dot{q}}{H}
\end{equation}
\begin{equation}
s = \frac{(r-1)}{3q-\frac{3}{2}}
\end{equation}
The expression of $r$ and $s$ for our cosmological model is
\begin{equation}
r = -1 + \frac{3\alpha(1-\Omega_{d})(1+2(\alpha-2\Omega_{d}))}{[2\alpha-\Omega_{d}(3\alpha-2)]^{2}} + \frac{3(3\alpha-2)(2\alpha-1)(\alpha-2)(1-\Omega_{d})\Omega_{d}}{[2\alpha-\Omega_{d}(3\alpha-2)]^{3}}
\end{equation}
\begin{equation}
s = \frac{\frac{3\alpha(1-\Omega_{d})(1+2(\alpha-2\Omega_{d}))}{[2\alpha-\Omega_{d}(3\alpha-2)]^{2}} + \frac{3(3\alpha-2)(2\alpha-1)(\alpha-2)(1-\Omega_{d})\Omega_{d}}{[2\alpha-\Omega_{d}(3\alpha-2)]^{3}}}{\frac{9\Omega_{d}(\alpha-2)+18\alpha}{2(2\alpha-\Omega_{d}(3\alpha-2))}}
\end{equation}
The trajectory of the ${r,s}$ statefinder parameters depicted in $Fig.11$ offers valuable insights into the dynamics of cosmic expansion described by the model. The current values of the statefinder parameters are $r_{0} = 0.74$ and $s_{0}=0.07$, placing them near the $\Lambda$CDM reference point $(1,0)$. The evolution of the trajectory originates from a region associated with earlier cosmic epochs, characterized by higher values of $r$ and positive $s$, and gradually approaches the $\Lambda$CDM point as the Universe evolves. This indicates a transition from a quintessence-like phase toward a behavior that closely resembles a cosmological constant.
The condition $r <1$ and $s >0$ in the present epoch implies that the dark energy component in the model exhibits mild dynamical features, distinguishing it from a strictly constant $\Lambda$, while still driving accelerated expansion. These results highlight the model's capacity to emulate $\Lambda$CDM behavior at low redshifts, thereby aligning well with contemporary cosmological observations.
\section{Conclusion} 
The scientific community's persistent curiosity about modern cosmological phenomena drives exploration of the Universe beyond conventional gravity models. This paper introduces a novel holographic dark-energy model, termed Fractional Holographic Dark Energy, which integrates concepts from fractional calculus and quantum mechanics into cosmology. A distinctive feature of this model is its new energy density formulation, which converges to the traditional holographic dark energy density as the parameter $\alpha$ approaches 2, where $\alpha$ reflects the characteristics of fractional mechanics. Specifically, we derived the energy density from the holographic inequality, incorporating corrections to the Bekenstein-Hawking entropy via the fractional Wheeler-De Witt equation. This yields a generalized energy density that extends the conventional holographic dark energy framework, offering new avenues for exploration. Using the Hubble-horizon cutoff, we then examined the cosmological evolution within this framework. We further applied this framework to Bianchi I cosmology, specifically focusing on the locally rotationally symmetric (LRS) model.\\
To constrain the model parameters, we employed a statistical Markov Chain Monte Carlo (MCMC) approach based on Bayesian inference. We then analyzed the results using various observational datasets, including Hubble, BAO, and Pantheon+Shoes(SNe). The constraint values for $H_{0}$ obtained are $67.688^{+1.246}_{-1.197}, \\67.74_{-1.25}^{+1.25}, 67.80^{+1.23}_{-1.23}$ and $67.73^{+1.23}_{-1.24}$ from the Hubble, Hubble + BAO, Hubble + SNe and Hubble + BAO + SNe datasets, respectively. From the Hubble, Hubble + BAO, Hubble + SNe, and Hubble + BAO + SNe datasets, the value of constraints viz. $\alpha$ and $\Omega_{m_{0}}$ are $(0.885^{+0.121}_{-0.149},0.264_{-0.018}^{+0.018}),\quad (0.87^{+0.14}_{-0.15},0.27_{-0.02}^{+0.02})$, $(0.86^{+0.14}_{-0.15},0.27_{-0.02}^{+0.02})$ and $(0.87^{+0.14}_{-0.14},0.27_{-0.02}^{+0.02})$ respectively, which are aligned with the well established observational data. Our analysis demonstrates that the deceleration parameter undergoes a transition from positive to negative values, indicating a shift from a decelerating to an accelerating phase of cosmic expansion. This behavior is consistent with observational evidence of the Universe’s late-time acceleration and offers a coherent description of its dynamic evolution. The equation of state (EoS) parameter $\omega_{d}$ exhibits a transition from quintessence-like behavior to values approaching $-1$, signifying a convergence toward $\Lambda$CDM-like dynamics at low redshifts. This transition reflects the model's ability to replicate the observed characteristics of dark energy in the late-time Universe. Furthermore, the statefinder diagnostic provides additional support for the model's consistency with observational data. The computed statefinder pair $(r,s)=(0.74,0.07)$ suggests that the model closely approximates $\Lambda$CDM characteristics within the considered redshift range. 
Our study demonstrates that the LRS Bianchi type-I cosmological model, incorporating fractional holographic dark energy, provides a robust framework for understanding the evolution of the Universe, consistent with observational data. Future investigations, leveraging more extensive datasets and refined theoretical models, will be vital for further validating and broadening the implications of these findings.

 $\textbf{Data Availability}$ No data are associated with the manuscript. \\
 $\textbf{Funding sources}$ This research did not receive any specific grant from funding agencies in the public, commercial, or non-profit sectors.\\

\end{document}